\documentclass[aps,prd,superscriptaddress,groupedaddress,nofootinbib]{revtex4}

\pdfoutput=1

\usepackage{amsmath}
\usepackage{graphicx}
\usepackage{color}
\usepackage{ulem}
\usepackage[strings]{underscore}
\usepackage{hyperref}
\usepackage{etoolbox}
\apptocmd{\thebibliography}{\raggedright}{}{}

\newcommand{\ud}{\mathrm{d}}

\newcommand{\uHo}{\mathrm{H_{0}}}

\newcommand{\cc}{\mathrm{C}}
\newcommand{\crad}{\mathrm{C_{\rm R}}}
\newcommand{\cmat}{\mathrm{C_{\rm M}}}
\newcommand{\OmegaR}{\Omega_{\rm R}}

\newcommand{\OmegaGW}{\Omega_{\rm gw}}

\newcommand{\uini}{\mathrm{ini}}
\newcommand{\tini}{t_\uini}

\newcommand{\pd}[2]{\left.\dfrac{\partial}{\partial #1}\right|_{#2}}
\newcommand{\calF}{n(t,\ell)}
\newcommand{\calP}{\mathcal{P}}
\newcommand{\calS}{\mathcal{S}}

\newcommand{\calJ}{\mathcal{J}}
\newcommand{\calJk}{\mathcal{J}_{\mathrm{k}}}
\newcommand{\calJc}{\mathcal{J}_{\mathrm{c}}}

\newcommand{\betak}{\beta_{\mathrm{k}}}
\newcommand{\betac}{\beta_{\mathrm{c}}}

\newcommand{\gammak}{\gamma_{\mathrm{k}}}
\newcommand{\gammac}{\gamma_{\mathrm{c}}}
\newcommand{\zf}{z_{\mathrm{friction}}}
\newcommand{\ff}{f_{\mathrm{friction}}}
\newcommand{\feq}{f_{\mathrm{eq}}}

\newcommand{\N}{\mathcal{N}}

\newcommand{\g}{g}
\newcommand{\lstar}{\ell_\star}
\newcommand{\tstar}{t_\star}
\newcommand{\ustar}{u_\star}
\newcommand{\lc}{\ell_{\rm c}}
\newcommand{\lk}{\ell_{\rm k}}

\newcommand{\zeq}{{z_{\rm eq}}}

\newcommand{\phiH}{\Phi_{\rm H}}
\newcommand{\ocas}{\omega_{\rm DGRB}}

\def\be{\begin{equation}}
\def\ee{\end{equation}}
\def\ba{\begin{eqnarray}}
\def\ea{\end{eqnarray}}
\def\nn{\nonumber}
\newcommand{\yo}{\ell_0}

\newcommand{\gammad}{\gamma_\ud}

\begin{document}

\title{Particle emission and gravitational radiation from cosmic strings: observational constraints}

\author{Pierre Auclair} \email{auclair@apc.univ-paris7.fr}
\affiliation{AstroParticule \& Cosmologie,
UMR 7164-CNRS, Universit\'e Denis Diderot-Paris 7,
CEA, Observatoire de Paris,
10 rue Alice Domon et L\'eonie
Duquet, F-75205 Paris Cedex 13, France}

\author{Dani\`ele A.~Steer} \email{steer@apc.univ-paris7.fr}
\affiliation{AstroParticule \& Cosmologie,
UMR 7164-CNRS, Universit\'e Denis Diderot-Paris 7,
CEA, Observatoire de Paris,
10 rue Alice Domon et L\'eonie
Duquet, F-75205 Paris Cedex 13, France}

\author{Tanmay Vachaspati} \email{tvachasp@asu.edu}
\affiliation{Physics Department, Arizona State University,
Tempe, AZ 85287, USA}

\begin{abstract}
We account for particle emission and gravitational radiation from cosmic string loops
to determine their effect on the loop distribution and observational signatures of strings.
The effect of particle emission is that the number density of loops no longer scales. This
results in a high frequency cutoff on the stochastic gravitational wave background, but we show that the expected
cutoff is outside the range of current and planned detectors. Particle emission from string
loops also produces a diffuse gamma ray background that is sensitive to the presence of
kinks and cusps on the loops.
However, both for kinks and cusps, and with mild assumptions about particle physics interactions,
current diffuse gamma-ray background observations do not constrain $G\mu$.

\end{abstract}

\maketitle

\section{Introduction}

Most often the dynamics of local cosmic strings formed in a phase transition in the early universe (see  \cite{ViSh,Us,TomMark} for reviews) is described by the Nambu-Goto (NG) action.  This approximation is valid when the microscopic width  of the string
\be
w\sim \mu^{-1/2} \sim 1/\eta
\label{wdef}
\ee
(with $\mu$ the string tension and $\eta$ the energy scale of the phase transition), is very small relative to its characteristic macroscopic size $\ell$ --- a situation which is well satisfied in the early universe.
Closed loops of NG strings loose energy slowly by radiating gravitational waves, and as a result NG string networks contain numerous loops whose decay generate a stochastic gravitational wave background (SGWB) ranging over a wide range of frequencies \cite{ViSh}. Depending on the details of the particular cosmic string model, the corresponding constraints on the dimensionless string tension $G \mu$ from the SGWB are $G\mu \lesssim 10^{-7}$ at LIGO-Virgo frequencies \cite{LigoStrings}, $G\mu \lesssim 10^{-11} $ at Pulsar  frequencies \cite{Blanco-Pillado:2017rnf}, whereas at LISA frequencies one expects to reach $G\mu \lesssim 10^{-17}$ \cite{LisaStrings}.

On the other hand, at a more fundamental level, cosmic strings are topological solutions of field theories.  Their dynamics can therefore also be studied by solving the field theory equations of motions. In studies of large scale field theory string networks \cite{Vincent:1997cx,Hindmarsh:2008dw,Lizarraga:2016onn,Hindmarsh:2017qff}, loops are observed to decay directly into particles and gauge boson radiation on a short time scale of order of the loop length.  Hence, field theory string network simulations predict very different observational consequences --- in particular no SGWB from loops.

Since field theory and Nambu-Goto strings in principle describe the same physics, and hence lead to
the same observational consequences,
this is an unhappy situation.
Based on high resolution field theory simulations, a possible answer to this long-standing conundrum was proposed in \cite{TVrecent}. In particular, for a loop of length $\ell$ containing {\it kinks},  a new characteristic length scale $\yo=\lk$ was identified, and it was shown that if  $\ell \gtrsim \lk$ gravitational wave emission is the dominant decay mode, whereas for smaller loops $\ell \lesssim \lk$ particle radiation is the primary channel for energy loss. That is,
\begin{equation}
\frac{d\ell}{dt}  = \begin{cases}
    -\gammad , & \ell \gg \lk \\
    -\gammad \frac{\lk}{ \ell} , & \ell \ll \lk,
\end{cases}
\label{kink}
\end{equation}
where
\be
\gammad \equiv \Gamma G \mu
\nn
\ee
with
$\Gamma \sim 50$ the standard constant describing gravitational radiation from cosmic string loops \cite{Vachaspati:1984gt,Burden:1985md,Garfinkle:1987yw,Blanco-Pillado:2017oxo}. Notice that Nambu Goto strings correspond to $\lk \rightarrow 0$; and if particle radiation is dominant for all loops,
$\lk \rightarrow \infty$.
In practise $\lk$ is neither of these two limiting values, and in \cite{TVrecent} was estimated (for a given class of loops with kinks) to be given by
\be
\lk \sim \betak \frac{w}{\Gamma G\mu}
\label{Est}
\ee
where $w$ is the width of the string, Eq.~(\ref{wdef}), and the constant $\betak\sim {\cal{O}}(1)$.

If a loop contains {\it cusps}, then one expects the above to be modified to \cite{BlancoPillado:1998bv,Olum:1998ag} 
\begin{equation}
       \frac{d\ell}{dt}  = \begin{cases}
           -\gammad , &  \ell \gg \lc \\
           -\gammad \sqrt{\frac{\lc }{ \ell} }, &  \ell \ll \lc
    \end{cases}
    \label{cusp}
    \end{equation}
where
\be
\lc
 \sim \betac \frac{w}{(\Gamma G\mu)^{2}}
 \label{lcdef}
\ee
with $\betac\sim {\cal{O}}(1)$.

The aim of this paper is to determine the observational effects --- and corresponding constraints on $G\mu$ --- of a finite, fixed, value of $\lk$ or $\lc$.  A first immediate consequence of the presence of the fixed scale  is that the distribution of loops $n(\ell,t)$, with $n(\ell,t)d\ell$ the number density of loops with length between $\ell$ and $\ell+d\ell$ at time $t$, will no longer be scaling. That is, contrary to the situation for NG strings, the loop distribution will depend explicitly on $t$ as well as the dimensionless variable $\gamma=\ell/t$.
We determine this non-scaling loop distribution $n(\gamma,t)$ in section \ref{sec:nlt}, taking into account exactly (and for the first time) the backreaction of particle emission on the loop distribution.

We then study the consequence of the non-scaling distribution of non-self intersecting loops on the stochastic GW background, determining the fraction of the critical density in GWs per logarithmic interval of frequency,
\be\label{eqn:theone}
\mathrm{\Omega}_{\rm gw}(t_0,f) = \frac{8 \pi G}{3\uHo^2 }~f ~ \frac{\ud\rho_{\rm gw}}{\ud f} (t_0, f)\,,
\ee
where $\uHo$ is the Hubble parameter, and
the $d\rho_{gw}/df$ factor
is the energy density in gravitational waves per unit frequency $f$ observed today (at $t=t_0$).  A scaling distribution of NG loops gives a spectrum which is flat at high frequencies \cite{ViSh}; we
will show below that a consequence of
the non-scaling of the loop distribution is the introduction of a characteristic frequency $f_*$, with $\Omega(f>f_*)\rightarrow 0$.
The precise value of $f_*$ depends on $\lk$ or $\lc$,
as well as $G\mu$. For cusps and kinks with $\lc$ and $\lk$ given respectively by Eqs.~(\ref{kink}) and (\ref{cusp}), the characteristic frequency $f_*$ is outside the LIGO and LISA band provided $G\mu \gtrsim 10^{-17}$, and so in this case the new cutoff will only be relevant for very light strings but for which the
amplitude of the signal is below the observational thresholds of planned gravitational wave detectors.

In section \ref{sec:particle} we turn to particle physics signatures.  At lower string tensions $G\mu$, the gravitational signatures of strings weaken, while the particle physics ones are expected to increase.  Following \cite{Sigl}, we focus on so-called ``top down'' models for production of ultra-high energy cosmic rays
in which heavy particles, namely the quanta of massive gauge and Higgs field of the underlying (local) field theory trapped inside the string, decay to give ultra-high energy protons and gamma rays.  We focus on the diffuse gamma ray flux which at GeV scales is constrained by Fermi-Lat \cite{FL}.
However, taking into account backreaction of the emitted particles on the loop distribution
we find that current gamma ray observations do not lead to significant constraints.
(Early studies on the production of cosmic rays assumed NG strings and particle
emission rates that were based on dynamics without taking backreaction into account.
See Refs.~\cite{Bhattacharjee:1989vu,MacGibbon:1989kk,MacGibbon:1992ug,Brandenberger:1993hw,Cui:2008bd} 
and~\cite{Sigl} for a review.
Other work has focused on strings with condensates, e.g.~\cite{Mota:2014uka,Vachaspati:2009kq,Peter:2013jj},
or strings coupled to other fields such as Kaluza-Klein or dilaton fields \cite{Dufaux,Damour:1996pv}.)

This paper is organised as follows.  In section \ref{sec:nlt} we determine the effect of an $\ell$-dependent energy loss
\begin{equation}
        \dfrac{\ud \ell}{\ud t} = - \gammad \calJ(\ell),
        \label{new}
    \end{equation}
on the loop distribution $n(\ell,t)$. The function $\calJ(\ell)$
will initially be left arbitrary. Specific cases corresponding to (i) NG loops with  $\calJ=1$; (ii) loops with kinks, see Eq.~(\ref{kink}), and (iii) loops with cusps, see Eq.~(\ref{cusp}) are studied in subsections \ref{ss:NG}-\ref{ss:c}.  Given the loop distribution, we then use it to calculate the SGWB in section \ref{sec:SGWB}, and the predicted diffuse gamma ray flux in \ref{sec:particle}.  We conclude in section \ref{sec:conc} by discussing the resulting experimental constraints on $G\mu$.

\section{The loop distribution}
\label{sec:nlt}

All observational consequences of string loops depend on $n(t,\ell) \ud \ell$, the number density of non self-intersecting loops with length between $\ell$ and $\ell+\ud\ell$ at time $t$. In this section we calculate $n(t,\ell)$ given (\ref{new}), that is we take into account the backreaction of the emitted particles on the loop distribution. As noted in the introduction, the existence of the fixed scale $\lk$ or $\lc$ means that the loop distribution will no longer  scale, that it will no longer be a function
of the dimensionless variable $\gamma \equiv \ell/t$.

\subsection{Boltzmann equation and general solution}

The loop distribution satisfies a Boltzmann equation which, taking into account the $\ell$-dependence of $\dot{\ell}$ (that is the flux of loops in $\ell$-space), is given by \cite{Copeland:1998na}
\begin{equation}
        \pd{t}{\ell} \left(a^3 \calF\right) + \pd{\ell}{t}\left(\dfrac{\ud \ell}{\ud t}a^3 \calF\right) = a^3 \calP
        \label{Boltzmann}
    \end{equation}
where $a(t)$ is the cosmic scale-factor, and the
loop production function (LPF)  $\calP(t,\ell)$ is the rate at which loops of length $\ell$ are formed at time $t$ by being chopped of the infinite string network.
On substituting (\ref{new}) into Eq.~(\ref{Boltzmann}) and multiplying each side of the equation
by $\calJ(\ell)$, one obtains
\begin{equation}
        \dfrac{1}{\gammad}\pd{t}{\ell} \g(t,\ell) - \calJ(\ell) \pd{\ell}{t}\g(t,\ell) = a^3 \calJ(\ell) \calP(t,\ell),
        \label{eq:this}
    \end{equation}
where
\begin{equation}
    g(t,\ell) \equiv \gammad \calJ(\ell) a^3(t) n(t,\ell).
    \label{eq:gdef}
\end{equation}

In order to solve (\ref{eq:this}), we first change variables from $(t,\ell)$ to
    \begin{equation}
        \tau \equiv \gammad t ~,~ \qquad \xi \equiv \int \dfrac{\ud \ell}{\calJ(\ell)}.
        \label{change1}
    \end{equation}
Notice from (\ref{new}) and (\ref{change1}) that for a loop formed at time $t_i$ with length $\ell_i$, its length at time $t$ satisfies
\be
\xi(\ell) + \gammad t = \xi(\ell_i) + \gammad t_i.
\label{physloop}
\ee
In terms of these variables Eq.~(\ref{eq:this}) reduces to a wave equation with a source term
    \begin{equation}
        \pd{\tau}{\xi} \g(\tau,\xi) - \pd{\xi}{\tau}\g(\tau,\xi) = \calS(\tau,\xi),
    \end{equation}
where
\be
\calS(\tau,\xi) = a^3(\tau)\calJ(\xi) \calP(\tau,\xi).
\nn
\ee
We now introduce the lightcone variables
    \begin{equation}
        2 u \equiv \tau - \xi ~,~ \qquad 2v \equiv \tau + \xi,
        \label{change2}
    \end{equation}
so that the evolution equation simply becomes
    \begin{equation}
        \pd{u}{v} g(u,v) = \calS(u,v),
        \label{eq:m}
    \end{equation}
which is straightforward to integrate. In the following we neglect any initial loop distribution at initial time $\tini$ (since this is rapidly diluted by the expansion of the universe), so that the general solution of (\ref{eq:m}), and hence the original Boltzmann equation Eq.~(\ref{Boltzmann}), is
    \begin{equation}
        g(u,v) = \int_{-v}^{u} \ud u' S(u',v).
        \label{eq:solution}
    \end{equation}
Finally one can convert back to the original variables $n(\ell,t)$ using (\ref{eq:gdef}) to find
    \begin{equation}
        n(t,\ell) = \frac{1}{\gammad \calJ(\ell)
        a^3(t)}\int_{-v(t,\ell)}^{u(t,\ell)} \ud u' \ a^3\big(u',v(t,\ell)\big) \calJ(u',v(t,\ell)) \calP(u',v(t,\ell))
    \end{equation}
where $v(t,\ell)$ is obtained from Eqs.~(\ref{change1}) and (\ref{change2}).
Notice that $\calJ$ appears in two places: as an overall factor in the denominator, as well as in the integrand.

\subsection{Solution for a $\delta$-function loop production function}
\label{subsec:delta}

We now assume that all loops are chopped off the infinite string network with
length $\alpha t$ at time $t$.  This assumption, which has often been used in the literature, will lead to analytic expressions. The value $\alpha \sim 0.1$ is suggested by the NG simulations of \cite{BlancoPillado:2011dq,Blanco-Pillado:2013qja},
particularly in the radiation era. However, one should note that other simulations \cite{Ringeval:2005kr} are consistent with power-law loop productions functions~\cite{Auclair:2019zoz,Lorenz:2010sm}, which have also been predicted analytically \cite{Polchinski:2006ee,Polchinski:2007rg,Dubath:2007mf}. These will be
considered elsewhere.  Since $\alpha t \gg (\lk,\lc)$ for $\alpha\sim0.1$, we expect that particle radiation from infinite strings will not affect the (horizon-size) production of loops from the scaling infinite string network, and hence we consider a loop production function of the form
    \begin{equation}
        \calP(t,\ell) = \cc t^{-5} \delta\left(\dfrac{\ell}{t} - \alpha\right)
    \end{equation}
 where the constant $\cc$, which takes different values in the radiation and matter eras, will be specified below.
  Substituting into \eqref{eq:solution}, assuming $a\propto t^{\nu}$, (with $\nu=1/2$ in the radiation era, and $\nu=2/3$ in the matter era) gives
    \begin{equation*}
        g(u,v) =\cc \int^u_{-v} \ud u' ~ \calJ[\ell(u',v)] t(u',v)^{-5} a[t(u',v)]^3
        \delta\left[ \dfrac{\ell(u',v)}{t(u',v)}-\alpha\right].
        \label{eq:mess}
    \end{equation*}

In order to evaluate this integral, in which $v=v(t,\ell)$ is {\it fixed}, let us denote the argument of the $\delta$-function by
\be
y \equiv \frac{\ell(u',v)}{t(u',v)}-\alpha .
\nn
\ee
For the given $v$, the argument vanishes ($y=0$) for some $u'(v)$, that we will denote $u_\star(v)$ and which therefore satisfies
 \begin{equation}
        \ell(\ustar,v) = \alpha t(\ustar,v).
        \label{tosolve}
    \end{equation}
Let us rewrite this more simply as $\lstar = \alpha \tstar$ where $\lstar \equiv \ell(\ustar,v)=\lstar(v)$ and  $\tstar \equiv t(\ustar,v) = \tstar(v)$.
Now, from the $v$ equation in \eqref{change2}, one has $2 v = \gammad t_\star(v) + \xi (\ell_\star(v))$. Furthermore --- since our final goal is to write the loop distribution in terms of $(t,\ell)$ (rather than $v$) --- we note from the same equation that $v$ is related to $(t,\ell)$ by $2v = \gammad t + \xi(\ell)$. Thus $\tstar(t,\ell)$, which will be required below, is the solution of
  \begin{equation}
        \gammad \tstar + \xi(\alpha \tstar) = \gammad t + \xi(\ell),
        \label{this}
    \end{equation}
which physically is simply relating the length of the loop $\alpha \tstar$ at its formation time $\tstar$, with its length $\ell$ at time $t$, see Eq.~(\ref{physloop}).

The final step needed to evaluate the integral in Eq.~(\ref{eq:mess}) is
the Jacobian of the transformation from $u'$ to $y$ which, on using (\ref{change2}), is given by
  \begin{equation*}
        \pd{u'}{v}\left(y(u',v)\right) = - \dfrac{\gammad \calJ(\ell(u',v)) t(u',v) + \ell(u',v)}{\gammad t(u',v)^2}.
    \end{equation*}
Evaluating this at $u'=\ustar$ and using $\ell_\star=\alpha t_\star$ gives
 \begin{equation*}
        \pd{u}{v}\left(y(\ustar,v)\right) = - \dfrac{\gammad \calJ[\alpha \tstar(t,\ell)] + \alpha }{\gammad \tstar(t,\ell)}.
    \end{equation*}
Having now expressed all the relevant quantities in terms of $(t,\ell)$, one can combine the above results and use the definition of $g$ in terms of $n(t,\ell)$ in Eq.~(\ref{eq:gdef}) to find
    \begin{equation}
         t^4 \calF =  \cc \dfrac{1}{\calJ(\ell)} \dfrac{\calJ(\alpha \tstar)}{\alpha +\gammad\calJ(\alpha \tstar)} \left( \frac{\tstar}{t}\right)^{-4}  \left(\frac{a(\tstar)}{a(t)}\right)^3.
         \label{eq:general}
    \end{equation}

This equation, which is exact, is the central result of this section and gives the loop distribution for any form of energy loss
$\ud \ell / \ud t = - \gammad \calJ(\ell)$,
provided the loop production function is a $\delta$-function. It generalises and extends other approximate results which may be found in the literature.

For loops that are formed in a given era (either radiation or matter domination) and decay in the {\it same} era, the above solution reduces to
\ba
             t^4 \calF &=&  \cc \dfrac{1}{\calJ(\ell)} \dfrac{\calJ(\alpha \tstar)}{\alpha +\gammad\calJ(\alpha \tstar)} \left( \frac{\tstar}{t}\right)^{3\nu-4}  .
        \label{eq:same}
\ea
 In the matter era, however, there also exists a population of loops which were {\it formed} in the radiation era, where $\cc=\crad$, and decay in the matter era. Indeed, this population generally dominates over loops formed in the matter era.  From (\ref{eq:general}) one can find a general expression for the distribution at any redshift $z$, provided the loops were formed in the radiation era ($\nu=1/2$): it is given by
        \begin{equation}
            t^4 n(t,\ell) =
             \crad \dfrac{1}{\calJ(\ell)} \dfrac{\calJ(\alpha \tstar)}{\alpha +\gammad\calJ(\alpha \tstar)} \left( \frac{\tstar}{t}\right)^{-5/2}  (1+z(t))^3 \left(2\sqrt{\OmegaR} \uHo t\right)^{3/2}
\label{looprm}
        \end{equation}
This reduces to (\ref{eq:general}) in the radiation era, and has the correct scaling in the matter era.

In the following we use standard Planck cosmology with Hubble constant $\uHo=100 h {\rm  km/s/Mpc}$, $h=0.678$, $\Omega_M=0.308$, $\OmegaR=9.1476\times 10^{-5}$ and $\Omega_{\Lambda} = 1-\Omega_M-\OmegaR$ \cite{Aghanim:2018eyx}. We model the varying number of effective degrees of freedom in the radiation era through $H(z) = \uHo {\cal H}(z)$ with
${\cal H}(z) = \sqrt{\Omega_\Lambda + \Omega_M(1+z)^3 + \Omega_R {\cal{G}}(z)(1+z)^4 }$
where ${\cal G}(z)$ is directly related to the effective number of degrees of freedom $g_*(z)$ and the effective number of entropic degrees of freedom $g_S(z)$ by \cite{Binetruy:2012ze}
\be
{\cal G}(z) = \frac{g_*(z) g_{S}^{4/3}(0)}{g_*(0) g_{S}^{4/3}(z)}.
\ee
We model this by a piecewise constant function whose value changes at the QCD phase transition ($T=200$MeV), and at electron-positron annihilation ($T=200$keV):
\be
{\cal G}(z) = \left\{\begin{array}{ll} \displaystyle
    1 & {\rm for}\ z<10^9,\\
    0.83 & {\rm for}\
    10^9<z<2 \times10^{12}.\\
    0.39 & {\rm for} \ z>2 \times10^{12}
\end{array}\right.
\label{dof}
\ee

\section{Loop distributions for particle radiation from cusps and kinks}
\label{sec:LD}

Given a specific form of $\calJ(\ell)$, the loop distribution $n(\ell,t)$ is given by (\ref{eq:general}), where $\tstar(t,\ell)$ is obtained by solving (\ref{this}).  The existence or not of an {\it analytical} solution depends on the form of $\calJ(\ell)$. In this section we consider three cases:
\begin{enumerate}
\item {\it Nambu-Goto loops}: here $\dot{\ell}=-\gammad$ so that $\calJ=1$;
\item {\it Loops with kinks}:  The asymptotic behaviour of $\calJ(\ell)$ is given in Eq.~(\ref{kink}).  This can be captured, for instance, by
$\calJ_1=1+\lk/\ell$
or alternatively by
    \begin{align}
        \calJk &= \sqrt{1+\left(\dfrac{\lk}{\ell}\right)^2}.
        \label{Jk}
    \end{align}
This second form gives a simpler analytic expression for $\tstar$, and we work with it below.  (We have checked that the differences in predictions arising from the choice of $\calJ_1$ or $\calJk$ are negligible.)

\item {\it Loops with cusps}: Following Eq.~(\ref{cusp}), we take
         \begin{equation}
         \calJ_c = \left[1+\left(\dfrac{\lc}{\ell}\right)^{3/2}\right]^{1/3},
         \label{Jc}
     \end{equation}
which has the correct asymtotic behaviour and also leads to analytical expressions.
An alternative, and seemingly simpler,  form $\calJ = 1+\sqrt{{\lc}/{\ell}}$ does not give
analytical expressions for $n(t,\ell)$.
\end{enumerate}

We now determine the corresponding loop distribution in scaling units, namely in terms of the variables
\be
\gamma \equiv \frac{\ell}{t}, \qquad \gammak(t) \equiv \frac{\lk}{t}, \qquad \gammac(t)\equiv\frac{\lc}{t},
\label{thegammasdef}
\ee
and determine
\be
{\N}(t,\gamma)\equiv t^4 n(t,\gamma).
\label{calNdef}
\ee

\subsection{NG strings}
\label{ss:NG}

A first check is that the above formalism yields the well known, standard, loop distribution for NG strings ($\calJ=1$). Eq.~(\ref{change1}) yields $\xi=\ell$, and from Eq.~(\ref{this}) it follows that
\be
\frac{\tstar}{t} = \frac{\gamma+\gammad}{\alpha+\gammad}.
\nn
\ee
Hence from Eq.~(\ref{eq:same})
    \begin{equation}
      {\N}_{NG}(t,\gamma) =  \cc   \frac{(\alpha+\gammad)^{3(1-\nu)}}{(\gamma+\gammad)^{4-3\nu}},
        \label{eq:NG}
    \end{equation}
which is the standard scaling NG loop distribution  for a delta-function loop production function \cite{ViSh}. In the radiation/matter eras, and on the scales $\alpha \gg \gammad$ observed in simulations, comparison with the numerical results of \cite{BlancoPillado:2011dq,Blanco-Pillado:2013qja,Ringeval:2005kr} sets the value of $\cc$ to respectively
\ba
\crad \alpha^{3/2} &\simeq& 0.18 \qquad (\text{radiation era})
\nn
\\
\cmat \alpha &\simeq& 0.27\qquad (\text{matter era})
\nn
\ea
The scaling distribution Eq.~(\ref{eq:NG}) is shown in the black (solid) curve in figure \ref{fig:kink}, where we have taken $\alpha=0.1$, $\gammad=10^{-6}$ and $\nu=1/2$ (radiation era).

\subsection{Loops with kinks}
\label{ss:kink}

\begin{figure*}
\includegraphics[width=0.75\textwidth]{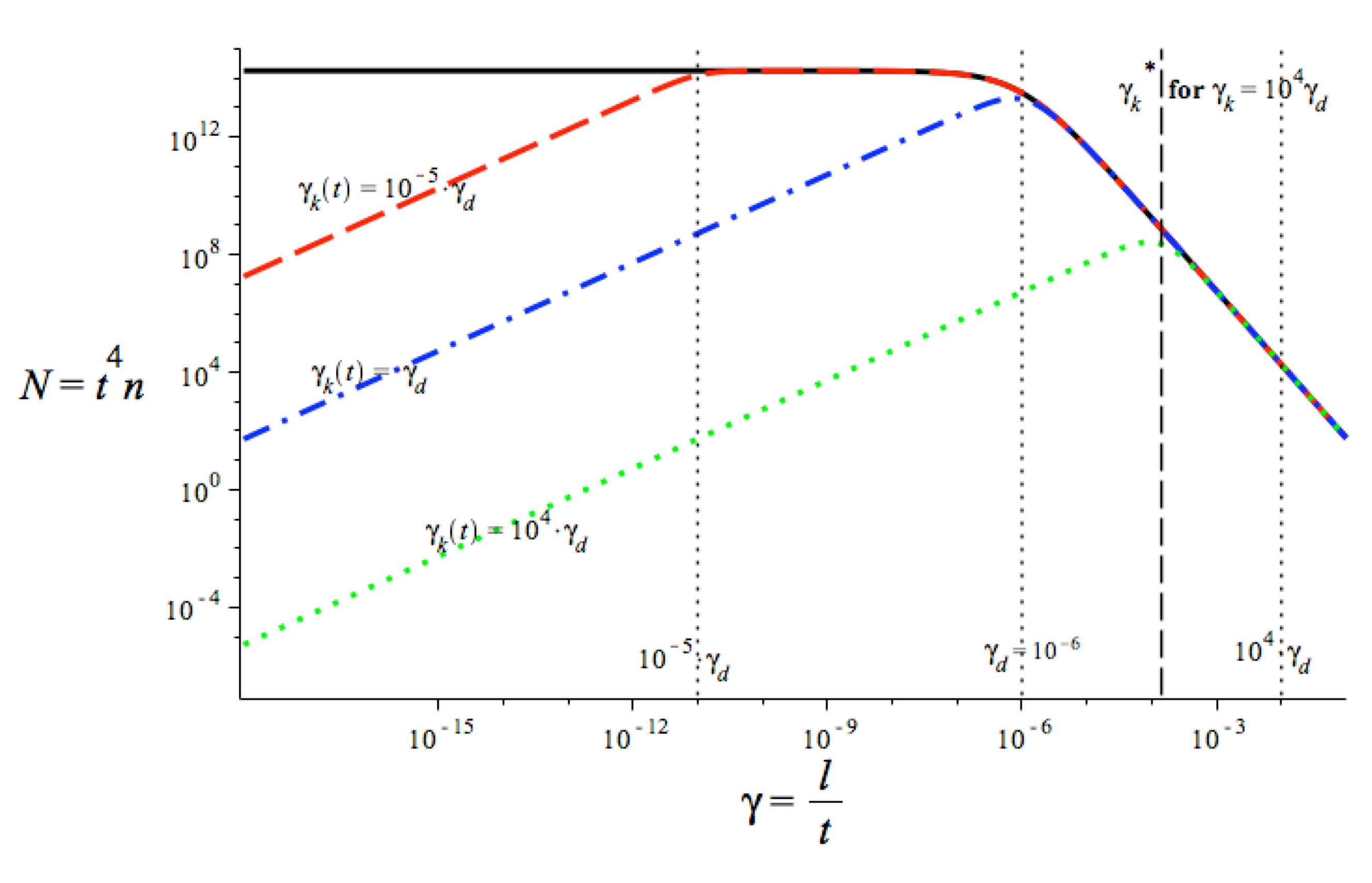}
\caption{
Loop distribution for kinks in the radiation era, with $\alpha=0.1$ and $\gammad = 10^{-6}$, and at several different epochs.  Black solid line: $\gammak=0$ ($t\rightarrow \infty$), the NG loop distribution. Red dash line: $\gammak(t)=10^{-5}\gammad$ (corresponding to $t=10^5 t_k$). Blue dot-dash line $\gammak(t)=\gammad$ (corresponding to $t=t_k$).  Green dotted line $\gammak(t)=10^4\gammad$ (corresponding $t=10^{-4}t_k$). }
\label{fig:kink}
\end{figure*}

From Eq.~(\ref{change1}), with $\calJ_k$ given Eq.~(\ref{Jk}), we now have $\xi(\ell) = \sqrt{\ell^2+\lk^2}$. Thus from Eq.~(\ref{this}), $\tstar$ satisfies a quadratic equation with solution
\be
 \frac{\tstar}{t} =\frac{ -\bar{\gamma}\left(\frac{\gammad}{\alpha}\right) + \sqrt{\bar{\gamma}^2-\gammak^2\left(1-\left(\frac{\gammad}{\alpha}\right)^2\right)}}{\alpha\left(1-\left(\frac{\gammad}{\alpha}\right)\right)}
 \ee
where $\gammak(t)$ is given in (\ref{thegammasdef}) and
\be
\bar{\gamma}(t,\gamma) \equiv  \gammad  + \sqrt{\gammak^2(t) +\gamma^2}
\ee
Since $\alpha \sim 0.1$ and $\gammad \equiv \Gamma G\mu \lesssim 10^{-6}$
(from cosmic microwave background constraints on cosmic strings \cite{Ade:2013xla})
in our analytical expressions below we ignore terms in $\gammad/\alpha$ so that
$\left(\alpha \tstar/t\right)^2=\bar{\gamma}^2-\gammak^2(t)$.
(This approximation was not used in our numerical calculations.)
Thus from Eq.~(\ref{eq:general}) we find, assuming $\alpha \gg \gammad$,
\be
{\N}(t,\gamma)= \cc \alpha^{3(1-\nu)}\left(\frac{\bar{\gamma}^2(t,\gamma)}{1+\gammak^2(t)/\gamma^2}\right)^{1/2}    \left(\bar{\gamma}^2(t,\gamma) - \gammak^2(t)\right)^\frac{3\nu-5}{2}  \qquad \qquad  {\text {where}} \; \gamma \leq \alpha,
\label{answerbis}
\ee
This distribution, in the radiation era, is plotted in
Fig.~\ref{fig:kink} for illustrative values of $\gammak(t)$, with $\gammad=10^{-6}$, $\alpha=0.1$.

The important qualitative and quantitative features to notice are the following:
\begin{itemize}
\item The existence of the fixed scale $\lk$ gives rise to a non-scaling distribution: $\N$ is explicitly $t$-dependent.
\item When $\gammak\rightarrow 0$, namely when $t\rightarrow \infty$,  Eq.~(\ref{answerbis}) reduces to the standard scaling NG loop distribution given in Eq.~(\ref{eq:NG}) (in the limit $\alpha \gg \gamma_d$).
\item For $\gamma \gg \gammak(t)$, the loop distribution is scaling since $\bar{\gamma}\sim \gamma + \gammad$, so that
\be
{\N}(t,\gamma) \simeq \cc \alpha^{3(1-\nu)}  (\gamma + \gammad)^{3\nu-4}.
\label{answersbig}
\ee
This behaviour is clear in Fig.~\ref{fig:kink} where for $\gamma \gg \gammak(t)$ the various curves coincide
with the NG curve. Hence for loops of these lengths, gravitational radiation is important but particle radiation plays no role. Furthermore
\begin{itemize}
\item  when $\gammad \gg \gamma \gg \gammak$, the distribution is {\it flat}, see figure \ref{fig:kink} dashed-red curve.
\item when $\gamma \gg (\gammad,\gammak)$ $\N$ drops off as $\gamma^{3\nu-4}$, as for NG loops, a dependence which is simply due to the expansion of the universe.
\end{itemize}
\item For $\gamma \ll \gammak(t)$, the distribution no-longer scales
because of particle radiation. Indeed $\bar{\gamma}\sim \gammak(t) + \gammad$ so that
\be
{\N} \simeq \cc \alpha^{3(1-\nu)} \gammad^\frac{3\nu-5}{2}
 \left( \frac{\gamma}{\gammak(t)}\right)
(2\gammak(t)+\gammad)^\frac{3\nu-5}{2} (\gammak(t)+\gammad).
\label{linear}
 \ee
This linear dependence on $\gamma$ for $\gamma \ll \gamma_k$ is visible in Fig.~\ref{fig:kink}. Notice that
\begin{itemize}
\item  when $\gammad \ll \gammak$, there is no plateau in the distribution, which goes from the linear behaviour Eq.(\ref{linear}) to the scaling behaviour Eq.~(\ref{answersbig}), at a value of $\gamma$ obtained by equating these two equations, namely
\be
\gammak^*(t)\simeq \sqrt{2\gammak\gammad}.
\nn
\ee
This is clearly visible in the green-dotted curve in Fig.~\ref{fig:kink}.
\end{itemize}

\end{itemize}
 When $\gammak(t)\ll \gammad$, an excellent approximation to the distribution is
\be
\N(\gamma,t) \simeq   \cc  \alpha^{3(1-\nu)} \dfrac{1}{\calJ(\gamma,t)}   (\gamma + \gammad)^{3\nu-4}.
        \label{eq:approx}
\ee
where, for the kinks considered here,
\be
\calJ(\gamma,t)=\sqrt{1+\left(\frac{\gammak(t)}{\gamma}\right)^2}.
\nn
\ee
On the other hand, when $\gammak(t)\geq \gammad$ the distribution changes behaviour, and for
$\gammak(t)\gg \gammad$
{\it its amplitude is significantly supressed due to particle emission}. Indeed when $\gamma=\gammak^*(t)$, which is at the maximum of $\N$ (see green curve, figure \ref{fig:kink}), ${\N}$ scales as $\gammak^{-(4-3\nu)/2}$ which decreases with increasing $\gammak$.  The equality $\gammad = \gammak(t) $ defines a {\it characteristic time}
$t_k$ by
\be
t_k \equiv \frac{\lk}{\gammad}.
\label{tkdef}
\ee
For $t\ll t_k$, particle emission is dominant, $\gammak(t)\geq \gammad$, and the distribution is supressed.
Using $\lk$ given by Eq.~(\ref{Est}),
\be
t_k
=\beta_k \frac{t_{pl}}{\Gamma^2 (G\mu)^{5/2}} \simeq \beta_k t_{eq} \left(\frac{2.5\times 10^{-24}}{G\mu}\right)^{5/2}
\nn
\ee
or in terms of redshift
\be
z_k \simeq z_{eq} \left(\frac{G\mu}{2.5\times 10^{-24}}\right)^{5/4} \frac{1}{\sqrt{\beta_k}}
\label{zkdef}
\ee
where
$z_{eq} \simeq \Omega_M/\OmegaR \sim 3367$.
The LH panel of Fig.~\ref{fig:kc} shows the loop distribution for different redshifts for $\lk$ given in Eq.~(\ref{Est}) and $\beta_k=1$. The effect of the supression of the loop distribution at  $z\gg z_k$ on the SGWB will be discussed in Sec.~\ref{sec:SGWB}.

\subsection{Loops with cusps}
\label{ss:c}
\begin{figure*}
      {\includegraphics{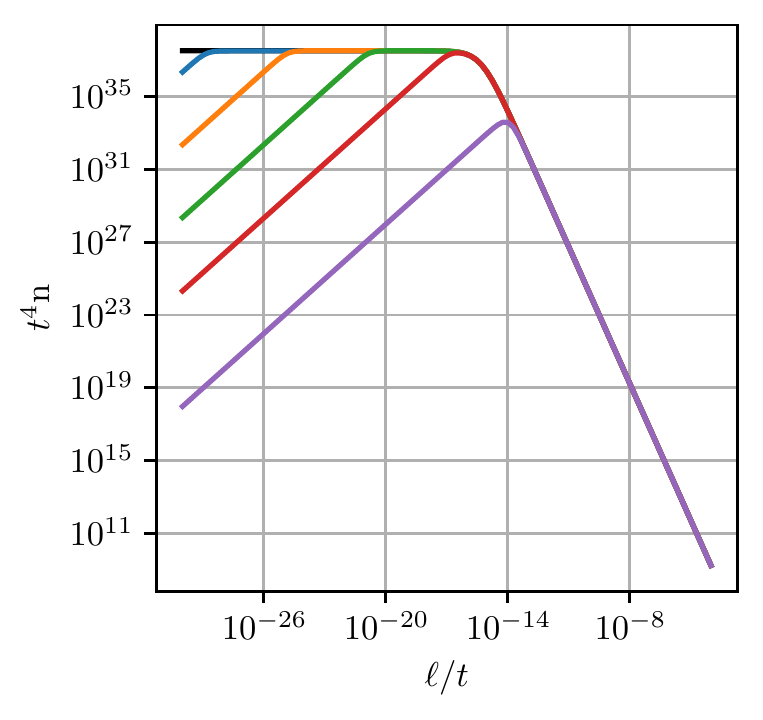}}
          {\includegraphics{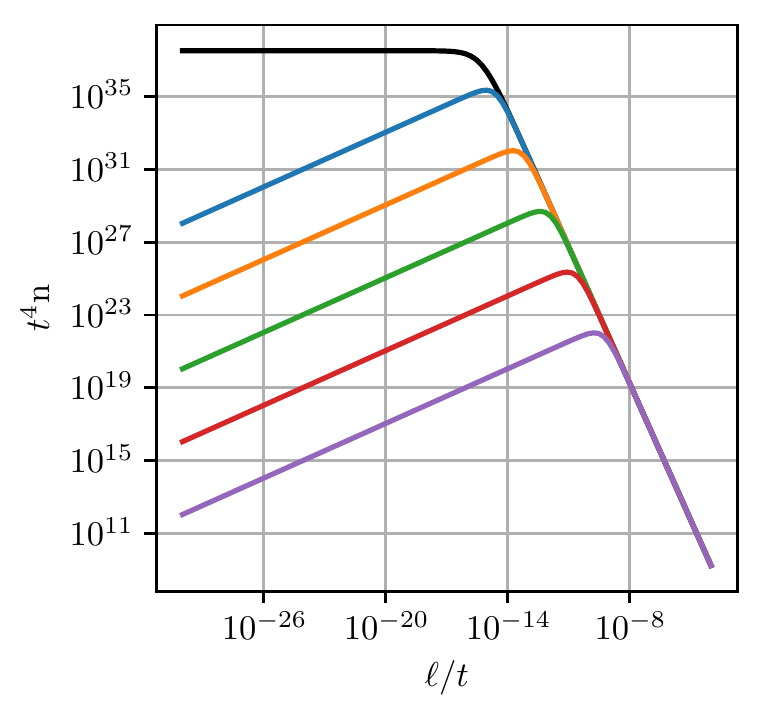}}
        \caption{
        Loop number density $\N=t^4 n$ for kinks [LH panel] and cusps [RH panel], for $G\mu=10^{-17}$.
        Thus $z_k\sim 10^{12}$ and $z_c \sim 10^{4}$. From bottom to top, the curves show
        snapshots of the loop distribution at redshifts $z=10^{13}, 10^{11}, 10^{9}, 10^{7}, 10^{5}$,
        and the black curve is the scaling loop distribution at $z\rightarrow 0$. The loop distributions are supressed for $z\gg z_k$ or $z\gg z_c$.
        }
        \label{fig:kc}
    \end{figure*}

For loops with cusps, where $\calJ=\calJc$ given in Eq.~(\ref{Jc}), the analysis is very similar. We only give the salient features.  As for kinks (see Eq.~\ref{tkdef}), one can define a characteristic time through $\gammad = \gammac(t)$, namely
\be
t_c \equiv \frac{\lc}{\gammad},
\label{tcdef}
\ee
and again, as for kinks, when $t \ll t_c$ the effects of particle radiation are more important and the loop distribution is supressed. For $\lc$ given in Eq.~(\ref{lcdef}), we have
\be
t_c
=\beta_c \frac{t_{pl}}{\Gamma^3 (G\mu)^{7/2}} \simeq \beta_c t_{eq} \left(\frac{4.6 \times 10^{-18}}{G\mu}\right)^{7/2}
\label{tcdefbis}
\ee
or in terms of redshift
\be
z_c \simeq z_{eq} \left(\frac{G\mu}{4.6 \times10^{-18}}\right)^{7/4} \frac{1}{\sqrt{\beta_c}}.
\label{zcdef}
\ee
For the relevant range, namely $G\mu < 10^{-6}$, we have $z_c < z_k$ and hence the observational consequences of cusps, both on the SGWB and the diffuse Gamma-ray background, are expected to be more significant than those of kinks --- since, as discussed above, the loop distribution is suppressed when $z<(z_c,z_k)$, see Fig.~\ref{fig:kc}.

The explicit $\gamma$-dependence of the distribution is the following.
First, substituting $\calJc$ in the definition of $\xi(\gamma)$ and $t_*$,
Eqs.(\ref{change1}) and (\ref{this}) respectively, we find
       \begin{align*}
            \xi(\ell) &= \left(\ell^{3/2}+\lc^{3/2}\right)^{2/3}, \\
             \left(\frac{\alpha \tstar}{t}\right)^{3/2} &= \left[\gammad + \left(\gamma^{3/2}+\gammac^{3/2}\right)^{2/3}\right]^{3/2} -\gammac^{3/2} \qquad \text{for} \; \alpha \gg \gammad.
        \end{align*}
It then follows from Eq.~(\ref{eq:same}) that the resulting distribution again scales for $\gamma \gg \gammac$ where it is given by Eq.~(\ref{answersbig}); and for $\gamma \ll \gammad$, ${\cal{N}} \propto \sqrt{\gamma}$. When $\gammac\gg \gammad$, we find
\begin{equation*}
            \cal{N} \propto \begin{cases}
                \gamma^{3\nu-4} \qquad \qquad (\gamma \gg \gammac^*) \\
                \sqrt{\gamma}\qquad \quad \qquad (\gamma \ll \gammac^*)
        \end{cases}
        \end{equation*}
where
\be
\gammac^* \simeq \left({\gammad \sqrt{\gammac}}\right)^{2/3}.
\nn
\ee

\section{The Stochastic Gravitational Wave Background}
\label{sec:SGWB}

The stochastic GW background $\mathrm{\Omega}_{\rm gw}(t_0,f)$ given in (\ref{eqn:theone}) is obtained by adding up the GW emission from all the loops throughout the whole history of the Universe which have contributed to frequency $f$.
Following the approach developed in \cite{Caldwell:1991jj,ViSh,Blanco-Pillado:2017oxo}
\begin{equation}
   \mathrm{\Omega}_{\rm gw}(\ln f) = \frac{8\pi G^2\mu^2 f}{3 \uHo^2}\sum_{j=1}^\infty C_j(f) P_j\,,
   \label{eqn:omega-method-1}
\end{equation}
where
\be\label{eqn:Cn}
C_j(f) = \frac{2j}{f^2 }\int_0^{\zf} \frac{dz}{H(z) (1+z)^6}~n\left(\frac{2j} {(1+z)f},t(z)\right)\,,
\ee
and $\zf$ is the redshift below which friction effects on the string dynamics become negligible \cite{ViSh}
\be
\zf \simeq
\zeq\, ( 4.4\times 10^{16})  \left(\frac{G\mu}{10^{-11}}\right) .
\label{zfr}
\ee
The $C_j$ depend on the loop distribution $n(\ell,t)$ through $n\left({2j} /{((1+z)f)},t(z)\right)$, whilst the $P_j$ are the ``average loop gravitational wave power-spectrum'', namely the power emitted in gravitational waves in the $j$th harmonic of the loop. By definition of $\Gamma$, these must be normalised to
$$
\Gamma = \sum_{j=1}^\infty P_j.
$$
For loops with kinks, $P_j  \propto j^{-5/3}$, whereas for loops with cusps $P_j  \propto j^{-4/3}$ ~\cite{Vachaspati:1984gt,Binetruy:2009vt,ViSh}.

\begin{figure*}
     \includegraphics{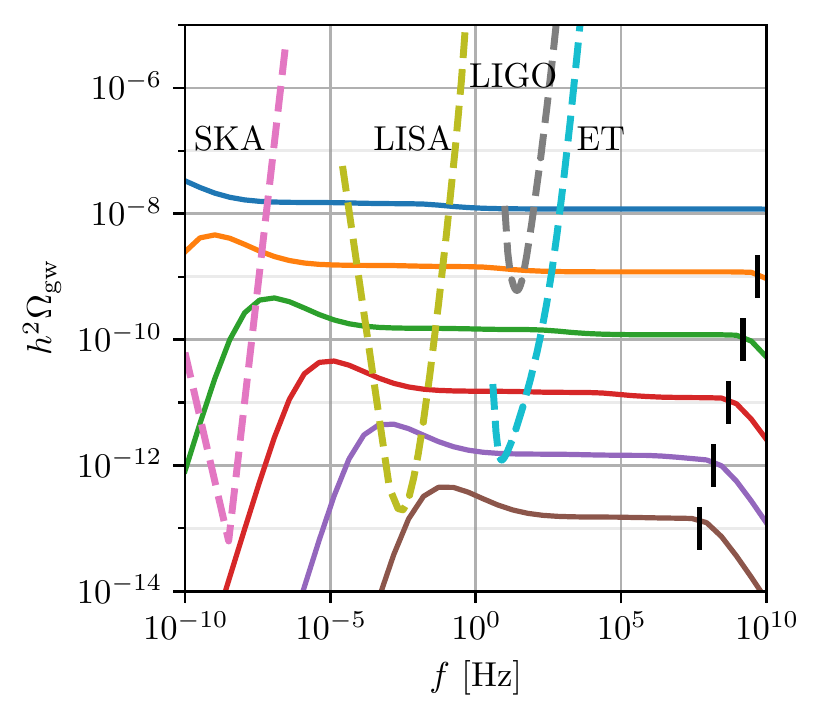}
      \includegraphics{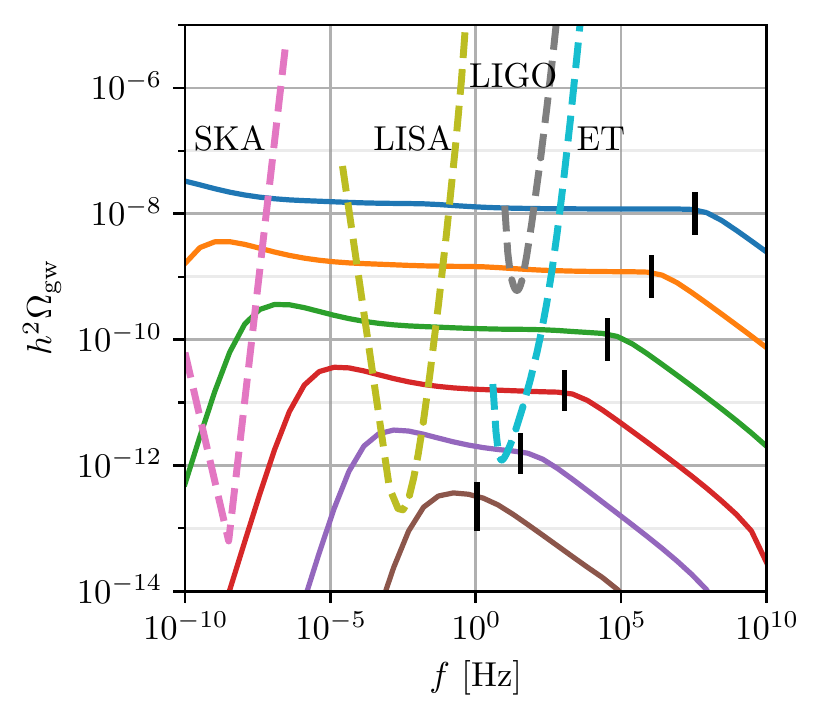}
      \caption{SBGW including the backreaction of particle emission on the loop distribution. LH panel: kinks on loops, RH panel: cusps on loop. The spectra are cutoff at high frequency, as indicated by the black vertical lines. $G\mu$ ranges from $10^{-17}$ (lower curve), through $10^{-15}$, $10^{-13}$,$10^{-11}$, $10^{-9}$ and $10^{-7}$ (upper curve). Also plotted are the power-law integrated sensitivity curves from SKA (pink dashed) \cite{Janssen:2014dka}, LISA (yellow dashed) \cite{Caprini:2019pxz}, adv-LIGO (grey dashed) \cite{TheLIGOScientific:2016dpb} and Einstein Telescope (blue dashed) \cite{Punturo:2010zz,Hild:2010id}.}
        \label{fig:one-scale-stochastic}
    \end{figure*}

As explained above, the effect of $\gammak$ and $\gammac$ on the loop distribution is particularly important at large redshifts $z>(z_c,z_k)$, and hence in the radiation era. Therefore we expect the effect of particle radiation to be visible in the high-frequency part of the spectrum.
This is indeed observed in Fig.~\ref{fig:one-scale-stochastic}, where the LH panel is for kinks with $\lk$ given in Eq.~(\ref{Est}) and $P_j \propto j^{-5/3}$; whereas the RH panel is for cusps with $\lc$ given in Eq.~(\ref{lcdef}) and $P_j \propto j^{-4/3}$.  As a result of the non-scaling loop distribution, the spectrum is no longer flat at high frequencies and, as expected, the effect is more significant for cusps than for kinks since $z_c < z_k$.

We can estimate the frequency above which the spectrum decays as follows.   In the radiation era
\begin{align}
        H(z) &= (1+z)^2  \sqrt{\OmegaR} \uHo \label{Hrad} \\
        t(z) &= \dfrac{1}{2(1+z)^2}\frac{1}{\sqrt{\OmegaR} \uHo} \label{trad}
    \end{align}
At high frequency, the lowest harmonic $j=1$ is expected to dominate \cite{ViSh}, so we set $P_j = \Gamma \delta_{j,1}$. Then using (\ref{Hrad}) and (\ref{trad}), Eq.~(\ref{eqn:omega-method-1}) simplifies to
\ba
   \mathrm{\Omega}_{\rm gw}(\ln f) &=& 2^4 \frac{16\pi (\Gamma G\mu)^2  }{3 \Gamma} \frac{\uHo}{ f } \OmegaR^{3/2} \int_{\zeq}^{\zf} {\ud z}~\N\left(\frac{2} {(1+z)f},t(z)\right)\,
   \nn
\\
   &\propto& \frac{\uHo}{ f }
   \left[ \int_{\zeq}^{z_{c,k}} {\ud z}~\N\left(\frac{2} {(1+z)f},t(z)\right) + \int_{z_{c,k}}^{\zf} {\ud z}~\N\left(\frac{2} {(1+z)f},t(z)\right) \right]\,.
   \nn
   \\
   &\simeq &\frac{\uHo}{ f } \int_{\zeq}^{z_{c,k}} {\ud z}~\N\left(\frac{2} {(1+z)f},t(z)\right) \,.
   \label{blimey}
\ea
Here, in going from the second to the third equality, we have used the fact that (i) for $G\mu \gtrsim 10^{-18}$,
which is relevant range for current and future GW detectors, $\zeq < (z_c,z_k) \ll \zf$ (see Eqs.~(\ref{zkdef}), (\ref{zcdef}) and (\ref{zfr})), and (ii) that the loop distribution above $z_{(c,k)}$ is subdominant, see e.g.~discussion above equation (\ref{tkdef}) in section \ref{ss:kink}. Using Eq.(\ref{trad}) as well as the approximation for the loop distribution for $z<z_k$ given in Eq.~(\ref{eq:approx}), it follows that for kinks
\ba
       [ \OmegaGW(\ln f)]_k
       &\propto& \int_{x_{\rm{eq}} }^{x_k} \left[1+\left(\dfrac{\lk x f^2}{8 \uHo \sqrt{\OmegaR}}\right)^2\right]^{-1/2}  \left(\gammad+x\right)^{-5/2} \ud x
       \label{rubbish}
\ea
where we have changed variable from $z$ to
\be
x = \frac{4}{f} (1+z) \uHo\sqrt{\OmegaR}
\nn
\ee
so that
\be
x_{\rm{eq}} = \frac{4}{f} (1+\zeq) \uHo\sqrt{\OmegaR}\, , \qquad x_k = \frac{4}{f} (1+z_k) \uHo\sqrt{\OmegaR}\, .
\nn
\ee

In order to understand the frequency dependence of $\OmegaGW$, let us initially
focus on the standard NG case, namely $\ell_k=0$.  (Here, the same change of
variable starting from the first line of Eq.~(\ref{blimey}) again yields Eq.~(\ref{rubbish})
but with upper bound replaced by $x_{\rm friction}={4} (1+\zf) \uHo\sqrt{\OmegaR}/f$).
Then Eq.~(\ref{rubbish}) gives
\be
 [ \OmegaGW(\ln f)]_{NG}\propto \frac{1}{\left(\frac{\feq}{f}+1\right)^{3/2}} -\frac{1}{\left(\frac{\ff}{f}+1\right)^{3/2}} ,
 \nn
 \ee
 where
\be
\feq = \frac{4 \uHo \sqrt{\OmegaR}(1+ \zeq)}{\gammad} \sim \frac{10^{-18}}{G\mu} {\rm s}^{-1}\, , \qquad  \ff = \frac{4 \uHo \sqrt{\OmegaR} (1+\zf)}{\gammad}  \sim 10^{10} {\rm s}^{-1},
\nn
\ee
and where in the last equality we have used Eq.~(\ref{zfr}).  At frequencies $f$ for which $\ff \gg f \gg \feq$ it follows that $ [ \OmegaGW(\ln f)]_{NG} \rightarrow$ constant
meaning that the spectrum is flat, which is the well known result for NG strings \cite{ViSh}.

For $\lk \neq 0$, the argument is altered because of the frequency dependence of the term in square brackets in Eq.~(\ref{rubbish}).  A further characteristic frequency now enters: this is can be obtained by combining the typical scales of the two terms in Eq.~(\ref{rubbish}). Namely, on one hand, from the first term (in square brackets) we have $\lk f^2\sim 8 \uHo \sqrt{\OmegaR}x^{-1}$; and on the other hand from the second (standard NG) term we have $x\sim \gammad$. Combining these yields the
characteristic frequency
   \begin{equation}
        f_k \sim \left( {\dfrac{8 \uHo \sqrt{\OmegaR}}{\lk \gammad}}\right)^{1/2} .
        \label{eq:dirac-cutoff}
    \end{equation}
For $f_k > f >f_{eq}$ the spectrum is still flat, as in the NG case. However, for $f>f_k$ it decays since the first term in square brackets in Eq.~(\ref{rubbish}) dominates. With $\lk$ given in Eq.~(\ref{Est}), $f_k \propto (G\mu)^{1/4}\betak^{-1/2}$, and this behaviour is clearly shown in  Fig.~\ref{fig:one-scale-stochastic} where $f_k$ is shown with a vertical black line for each value of $G\mu$ and we have assumed $\betak=1$.

For cusps the analysis proceeds identically with
    \begin{equation}
        f_c = \left( {\dfrac{8 \uHo \sqrt{\OmegaR}}{\lc \gammad}}\right)^{1/2} .
        \label{eq:dirac-cutoffc}
    \end{equation}
Now, on using $\lc$ defined in Eq.~(\ref{lcdef}), we have $f_c \propto (G\mu)^{3/4}\betac^{-1/2}$.  The spectrum of SGWB in this case is shown in the RH panel of Fig.~\ref{fig:one-scale-stochastic} where $f_c$ is shown with a vertical black line for each value of $G\mu$ and we have taken $\betac=1$.

As the figure shows, with $\betac=1$ and in the range of $G\mu$ of interest for GW detectors,
the decay of $\Omega_{\rm{GW}}$ for $f>f_c$  is {\it outside} the observational window of the
LIGO, LISA (and future ET) detectors.  In order to have $f_c \sim f_{\rm {LIGO}}$, one would require large
values of $\betac$ which are not expected.

\section{Emission of particles}
\label{sec:particle}

The loops we consider radiate not only GW but also particles. Indeed, for loops with kinks,
from Eq.~(\ref{kink})
\be
\left. \dot{\ell} \right|_{\rm particle} = -\gammad \frac{\lk}{\ell}
\label{rp}
\ee
The emitted particles are heavy and in the dark particle physics sector corresponding to the
fields that make up the string. We assume that there is some interaction of the dark sector
with the standard model sector. Then the emitted particle radiation will eventually decay,
and a significant fraction of the energy $f_{{\rm eff}} \sim 1$ will cascade down into $\gamma$-rays.
 Hence the string network will be constrained by the Diffuse Gamma-Ray bound measured at GeV scales by Fermi-Lat \cite{FL}. This bound is
\be
\ocas^{\rm obs} \lesssim 5.8 \times 10^{-7} \; {\rm eV}{\rm cm}^{-3},
\ee
where $\ocas$ is the total electromagnetic energy injected since the universe became
transparent to GeV $\gamma$ rays at $t_{\gamma} \simeq 10^{15}$s, see e.g.~\cite{Mota:2014uka}.

The rate per unit volume at which string loops lose energy into particles can be obtained by integrating (\ref{rp}) over the loop distribution $n(\ell,t)=t^{-4}\N(\gamma,t)$, namely
    \begin{equation}
        \phiH(t) = \mu \gammad {\lk} \int_0^{\alpha t} n(\ell,t)\frac{\ud\ell}{\ell} =
          \mu t^{-3}  {\gammad \gammak} \int_0^{\alpha} \  \frac{{\N}(\gamma',t)}{\gamma'} \ud \gamma'
    \end{equation}
The Diffuse Gamma Ray Background  (DGRB)  contribution is then given by (see e.g.~\cite{Mota:2014uka})
\ba
\ocas &=& f_{\rm eff }\int_{t_\gamma}^{t_0} \dfrac{{\rm \Phi_H}(t)}{(1+z)^4} \ud t
\nn
\\
&=& f_{\rm eff } \mu  {\gammad } \int_{t_\gamma}^{t_0}   \dfrac{\gammak(t)}{t^3 (1+z(t))^4}  \left[ \int_0^{\alpha} \  \frac{{\N}(\gamma',t)}{\gamma'} \ud \gamma' \right] \ud t
\nn
\\
&=&
  \Gamma(8.4\times10^{39})  f_{\rm eff } \left(\frac{G\mu}{c^4}\right)^2
  \int_{t_\gamma}^{t_0}   \dfrac{\gammak(t)}{t^3 (1+z(t))^4}  \left[ \int_0^{\alpha} \  \frac{{\N}(\gamma',t)}{\gamma'} \ud \gamma' \right] \ud t
  \qquad {\rm eV}{\rm cm}^{-3}
  \label{omgk}
\ea
where in the last line we have explicity put in factors of $c$ converted to physical units of ${\rm eV}/{\rm cm}^3$.
For cusps, one finds
\ba
\ocas &=& \Gamma(8.4\times10^{39})  f_{\rm eff } \left(\frac{G\mu}{c^4}\right)^2   \int_{t_\gamma}^{t_0}   \dfrac{\sqrt{\gammac(t)}}{t^3 (1+z(t))^4}  \left[ \int_0^{\alpha} \  \frac{{\N}(\gamma',t)}{\sqrt{\gamma'}} \ud \gamma' \right] \ud t \qquad {\rm eV}{\rm cm}^{-3}
  \label{omgc}
\ea

\begin{figure*}
       {\includegraphics[width=0.45\textwidth]{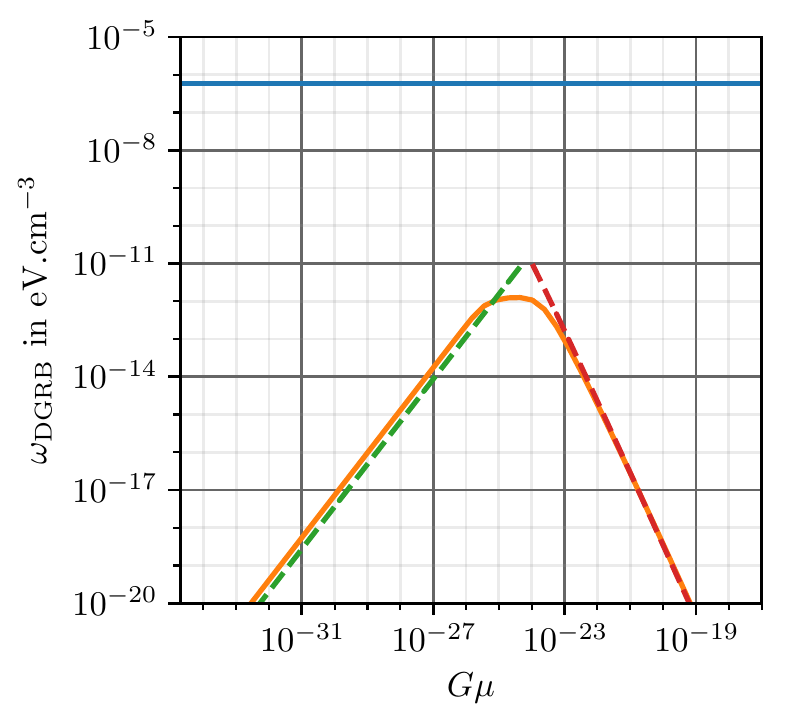}}
       {\includegraphics[width=0.45\textwidth]{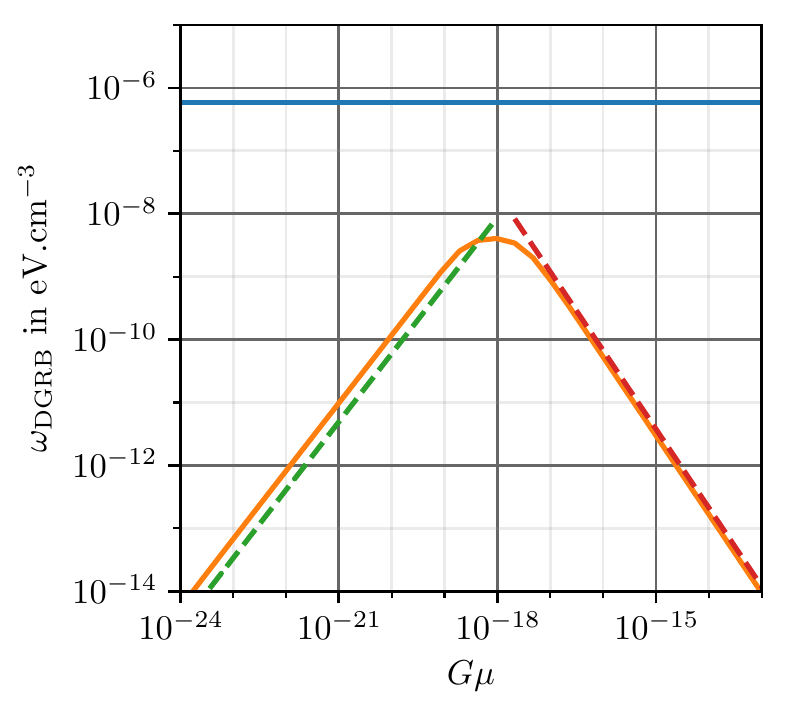}}
        \caption{
        Contribution of cosmic strings to the Diffuse Gamma-Ray Background.  The (blue) horizontal line is the experimental constraint from Fermi-LAT, while the (orange) line is the exact numerical calculation for kinks (LH panel) and cusps (RH panel).  On either side of the maxima, the slope and amplitude can be estimated using the results of previous sections.  Kinks: for low $G\mu$ the slope is $9/8$ (dashed-green line), and for high $G\mu$ it depends on $\mu^{-2}\log(\mu)$ (dashed-red line).
        Cusps: For low $G\mu$ the slope is $13/12$ (dashed-green line), and for high $G\mu$ it is $-5/4$  (dashed-red line).  The slightly different amplitude between the numerical calculation and the analytical one is because the latter assumes a matter dominated universe, and hence neglects effects of late time acceleration.}
        \label{fig:omg}
    \end{figure*}

In the matter dominated era, the loop distribution is dominated by those loops produced in
the radiation era but decay in the matter era: its general expression is given in Eq.~(\ref{looprm}),
and can be deduced straightforwardly from the results of subsections \ref{ss:kink} and \ref{ss:c}
for kinks and cusps respectively.  We have calculated (\ref{omgk}) and (\ref{omgc}) numerically,
 and the results are shown in
Fig.~\ref{fig:omg} for kinks [LH panel] and cusps [RH panel], together with the Fermi-Lat bound.
It is clear from this figure that particle radiation from loops containing
kinks and/or cusps, with $\lk$ and $\lc$ given in (\ref{Est}) and \eqref{lcdef},
are not constrained by the Fermi-lat data.

 The general shape of the spectra in Fig.~\ref{fig:omg} can be understood  from the
 results of section \ref{sec:nlt}. Let us focus on the case of cusps (for kinks the analysis is similar).
 First, we can determine the range of $G\mu$ for which the characteristic time $t_c$ defined in
  Eq.~(\ref{tcdef}) falls within the range of integration of (\ref{omgc}), namely
 \be
 t_\gamma \leq t_c \leq t_0  \, \qquad \Longleftrightarrow \qquad  10^{-19} \lesssim G\mu \lesssim 10^{-18}
 \nn
 \ee
 (we have assumed $\betac=1$ and, from Eq.~(\ref{tcdefbis}), $t=t_c$ implies
 $G\mu \sim 4.6\times 10^{-18} ( {t_{\rm eq}}/{t})^{2/7}$). This range of $G\mu$ defines the position
 of the maximum of the DGRB  in the RH panel of Fig.~\ref{fig:omg}. For lower $G\mu$, all times in
 the integration range are {\it smaller} than $t_c$.  As we have discussed in Sec.~\ref{ss:c}, in this
 case the loop distributions are {\it supressed} due to particle radiation: there are fewer loops, and
 hence fewer particles are emitted leading to a decrease in the DGRB. This is shown in
 Fig.~\ref{fig:omg}, and using the results of  Sec.~\ref{ss:c}, one can show that for
 $G\mu \ll 10^{-19}$, $\phiH(t)  \propto \mu^{2/3} \lc^{-1/6} (1+z)^3 t^{-4/3}$ leading to
\be
                \ocas  \propto \mu^{2/3} \lc^{-1/6} \propto (G\mu)^{13/12} \qquad (G\mu \ll 10^{-19}).
                \nn
\ee
On the other hand, for $G\mu \gg 10^{-18}$, all times in the integration range are {\it larger} than $t_c$. There is no supression of the loop distribution, since GR dominates over particle emission (see Sec.~\ref{sec:nlt}). But precisely because GR dominates, fewer particles are emitted, and hence we also have a decrease in the DGRB.   We now find that $\phiH(t)\propto  \gammad^{-1} \mu \sqrt{\lc} (1+z)^3 t^{-2}$ so that
\be
                \ocas \propto \sqrt{\lc} \propto (G\mu)^{-5/4}
\nn
\ee
which is the slope seen in Fig.~\ref{fig:omg}. For kinks the discussion is very similar, and the slopes are given in the caption of the figure. However, each kink event emits fewer particles, leading to a lower overall DGRB.

\section{Conclusion}
\label{sec:conc}

Cosmic string loops emit both particle and gravitational radiation. Particle emission is more important
for small loops, while gravitational emission dominates for large loops. In this work, we have
accounted for both types of radiation in the number density of loops and calculated the expected stochastic gravitational wave background and the diffuse gamma ray background from strings.
Our results show that the
number density of loops gets cutoff at small lengths due to particle radiation. The strength of
the cutoff depends on the detailed particle emission mechanism from strings -- if only kinks
are prevalent on strings, small loops are suppressed but not as much as in the case when
cusps are prevalent (see Fig.~\ref{fig:kc}). The cutoff in loop sizes implies that the stochastic
gravitational wave background will get cut off at high frequencies (see Fig.~\ref{fig:one-scale-stochastic}).
The high frequency cutoff does not affect current gravitational wave detection efforts but may
become important for future experiments.

Particle emission from strings can provide an important alternate observational signature
in the form of cosmic rays. Assuming that the particles emitted from strings
decay into standard model Higgs particles that then eventually cascade into gamma rays,
we can calculate the gamma ray background from strings. This background is below
current constraints in the case of both kinks and cusps.

It is important to evaluate more carefully the prevalence of kinks versus cusps on
cosmological string loops. In \cite{TVrecent}, particle radiation from a loop of a specific shape
was studied where the shape was dictated by general expectations for the behavior
of the cosmological string network. That particular loop only contained kinks. It would be
of interest to study other loop shapes that are likely to be produced from the network
and that contain cusps and to assess if the $1/\sqrt{\ell}$ dependence in \eqref{cusp} is
an accurate characterization of such loops over their lifetimes. It would also be interesting to study other loop production functions, particularly those of \cite{Polchinski:2006ee,Polchinski:2007rg,Dubath:2007mf} which predict a larger number of small loops relative to the situation studied in section \ref{subsec:delta}; hence one might expect a larger gamma ray background from strings in this case\footnote{Work in progress}.

\acknowledgements
We would like to thank Dimitri Semikoz and Ed Porter for useful discussions, and Mark Hindmarsh, Christophe Ringeval and G\'eraldine Servant for useful comments and questions on the first draft of this paper. PA thanks Nordita for hospitality whilst this work was in progress. DAS thanks Marc Ar\`ene, Simone Mastrogiovanni and Antoine Petiteau. TV thanks APC (Universit\'e Paris Diderot)
for hospitality through a visiting Professorship while this work was being done.
TV is supported by the U.S. Department of Energy, Office of High Energy Physics, under
Award No.~DE-SC0019470 at Arizona State University.

\noindent
\bibliography{refs}

\end{document}